\documentclass[aps,prl,showpacs,superscriptaddress,floatfix,twocolumn]{revtex4-1}
\usepackage{times}
\usepackage{graphicx}
\usepackage{amsmath}
\usepackage{amssymb}
\usepackage{subfigure}
\usepackage{color}
\usepackage{epstopdf}

\begin{document}

\title{Local Electronic Structure and Fano Interference in Tunneling into a Kondo Hole System}

\author{Jian-Xin Zhu}
\email[To whom correspondence should be addressed.\\]{jxzhu@lanl.gov}
\homepage{http://theory.lanl.gov}
\affiliation{Theoretical Division, Los Alamos National Laboratory,
Los Alamos, New Mexico 87545, USA}

\author{Jean-Pierre Julien}
\affiliation{Institut Neel CNRS and Universit\'{e} J. Fourier 25 Avenue des Martyrs, BP 166, F-38042
Grenoble Cedex 9, France}

\author{Y. Dubi}
\affiliation{School of Physics and Astronomy, Tel-Aviv University, Tel-Aviv, Israel }

\author{A. V. Balatsky}
\affiliation{Theoretical Division, Los Alamos National Laboratory,
Los Alamos, New Mexico 87545, USA}
\affiliation{Center for Integrated Nanotechnologies, Los Alamos National Laboratory, Los Alamos, New Mexico 87545, USA}

\begin{abstract}
Motivated by recent success of local electron tunneling into heavy fermion materials, we study the local electronic structure around a single Kondo hole in an Anderson lattice model and the Fano interference pattern relevant to STM experiments. 
Within the Gutzwiller method, we find that an intragap bound state exists
 in the heavy Fermi liquid regime. 
The energy position of the intragap bound state is dependent on the on-site potential scattering 
strength in the conduction and $f$-orbital channels. 
Within the same method, we derive a new $dI/dV$ formulation, which includes explicitly  the renormalization effect due to the $f$-electron correlation.
It is found that the Fano interference gives  asymmetric coherent peaks separated by the hybridization gap. The intragap peak structure has a Lorenzian shape, and the corresponding 
$dI/dV$ intensity depends on the energy location of the bound state. 
\end{abstract}
\maketitle

{\it Introduction.~} The intermetallic heavy fermion compounds 
based on either rare earth elements or on actinides~\cite{GRStewart84,HTsunetsugu97,GRStewart01} exhibit 
 many unusual properties like  heavy Fermi liquid (HFL),
magnetic ordering, quantum phase transitions and associated non-Fermi liquid, 
as well as unconventional superconductivity~\cite{PGegenwart07}.
In these materials, there
are two types of electrons: delocalized conduction electrons ($c$-electrons),
which derive from outer atomic orbitals, and strongly localized
$f$-electrons that singly occupy inner orbitals. It is well accepted that 
the interplay between the $c$-$f$ hybridization and the screened on-site 
Coulomb repulsion is the key to the above-mentioned  anomalous properties. 
However, the precise nature of the coherent Kondo state responsible 
for the formation of HFLs remains hotly debated~\cite{SShiba:2005,SBurdin:2000,YYang:2008,LZhu:2010}.  
Recent success of scanning tunneling microscopy (STM) experiments on heavy fermion 
materials~\cite{ARSchmidt:2010,PAynajian:2010,SErnst:2011} opens a new avenue toward understanding of these remarkable properties. Interest in the problem was also stimulated by the possibility to reveal the nature of electronic correlation effects by impurities and defects in heavy fermion materials as in unconventional superconductors including cuprates~\cite{AVBalatsky:2006} and iron 
pnictides~\cite{DZhang:2009,TZhou:2009,WTsai:2009} or selenides~\cite{JXZhu:2011}. 

Theoretical challenge in the interpretation of differential tunneling conductance as measured 
by STM in heavy-fermion lattice systems and around a single magnetic impurities or, more generally, disordered Kondo lattice systems, requires 
a proper treatment of electron correlation effects and quantum interference between the electrons
tunneling from the STM tip into the conduction band and into the magnetic $f$-electron states.
In the single Kondo impurity case, the line shape of $dI/dV$ has been described reasonably well with a phenomenological form, as first discussed by Fano~\cite{UFano:1961}. A microscopic understanding of this Fano line shape was recently provided~\cite{MMaltseva:2009,JFiggins:2010} in the single Kondo impurity case. In addition, the essence of a similar line shape observed on a clean crystal surface~\cite{ARSchmidt:2010,PAynajian:2010}, which has the lattice translational symmetry, can be captured 
within a similar microscopic approach~\cite{MMaltseva:2009,JFiggins:2010,PWolfle:2010,YDubi:2011}. However, there is still a significant question about the role of 
correlation effect in the Fano interference.
In one scenario~\cite{MMaltseva:2009}, the passage of an electron from the STM tip is accompanied by a simultaneous spin flip of the localized moments via cotunneling mechanism. When the local spin operator is represented in terms of pseudofermions, this cotunneling term is equivalent to an effective tunneling in the pseudofermion channel renormalized by a local boson condensation order parameter inherent to the Kondo lattice itself~\cite{MMaltseva:2009}.  In another scenario within the same Kondo lattice model~\cite{JFiggins:2010}, the renormalization factor is absent in the $dI/dV$ formula.
For both the single Kondo impurity problem and the Kondo lattice model, since the renormalization factor can be absorbed into the bare tunneling amplitude, this difference will not cause qualitative difference in analysis of STM data. However, for a Kondo hole or a more general disordered Kondo lattice problem, a logically consistent $dI/dV$ formulation is important. In connection with the STM measurements, the latter type of problems just began to attract more focused interest because the 
inhomogeneity may provide important insight into the complex electronic structure of heavy fermion materials. 

The Kondo hole problem was previously studied within an Anderson lattice model, where the $f$-electron self-energy for the pristine system was obtained from the second-order perturbation in Hubbard repulsion~\cite{RSollie:1991}. In the Kondo insulator (KI) regime, in which a hybridization gap is open at the Fermi energy, it was shown from the $f$-electron local density of states~\cite{RSollie:1991} that the Kondo hole introduced an intragap bound state. However, a very recent study of a single Kondo hole 
problem within the Kondo lattice model indicated that the existence of the intragap bound state depends on the nature of defects  in the HFL regime~\cite{JFiggins:2011}. This discrepancy suggests further theoretical studies are needed. In addition, 
existing studies~\cite{RSollie:1991,JFiggins:2011} are limited to local density of states and an analysis of Fano interference in the Kondo hole problem is still lacking. 
In this Letter, we study the local electronic structure around a single Kondo hole in an Anderson lattice model as well as the Fano interference pattern relevant to STM experiments. 
Within the Gutzwiller method, we are able to demonstrate that the intragap bound state exists regardless of whether the system is in the HFL or KI regime. 
The energy position of the intragap bound state is dependent on the on-site potential scattering 
strength in the conduction and $f$-orbital channels. 
The same Gutzwiller method enables us to include explicitly the renormalization effect due to the $f$-electron correlation into the $dI/dV$ formulation, which is conceptually consistent with the cotunneling mechanism~\cite{MMaltseva:2009}. It is found that the Fano interference does give rise to an asymmetric coherent peaks separated by the hybridization gap, while it tunes the intragap peak structure dependent on the energy location of the bound state, as shown in $dI/dV$. 

{\em  Kondo hole model and Gutzwiller formalism.~}
Our Hamiltonian for the single Kondo hole system is written as 
\begin{eqnarray}
H &=& - \sum_{ij,\sigma} (t_{ij}^{c} + \mu \delta_{ij})c_{i\sigma}^{\dagger} c_{j\sigma}  + \sum_{i,\sigma}  [V_{cf}  c_{i\sigma}^{\dagger} f_{i\sigma} + \text{H.c.}] \nonumber \\
&& + \sum_{i,\sigma} (\epsilon_{f}-\mu) f_{i\sigma}^{\dagger} f_{i\sigma}
+ \sum_{i} U_{f} n_{i\uparrow}^{f} n_{i\downarrow}^{f}  \nonumber \\
&&+ \sum_{\sigma} \epsilon_{c}^{I} c_{I\sigma}^{\dagger} c_{I\sigma} +   \sum_{\sigma} (\epsilon_{f}^{I} - \epsilon_{f}) f_{I\sigma}^{\dagger} f_{I\sigma}
- U_{f} n_{I\uparrow}^{f} n_{I\downarrow}^{f}
 \nonumber \\
&& +\sum_{\sigma}  [(V_{cf}^{I}-V_{cf})  c_{I\sigma}^{\dagger} f_{I\sigma} + \text{H.c.}]  \;.
\label{EQ:Hamil}
\end{eqnarray}
Here the operators $c_{i\sigma}^{\dagger}$ ($c_{i\sigma}$) create (annihilate) a conduction electron at site $\mathbf{r}_{i}$ with spin projection $\sigma$ while the operators $f_{i\sigma}^{\dagger}$ 
($f_{i\sigma}$) create (annihilate) a $f$-level electron at site $\mathbf{r}_{i}$ with spin projection $\sigma$. 
The number
operators for $c$ and $f$ orbitals with spin $\sigma$ are given by
$n_{i\sigma}^{c}=c_{i\sigma}^{\dagger}c_{i\sigma}$ and
$n_{i\sigma}^{f}=f_{i\sigma}^{\dagger}f_{i\sigma}$,
respectively. 
The quantity $t_{ij}^{c}$ is the hopping integral of the conduction electrons, and $\epsilon_{f}$ is the local $f$-orbital energy level on the magnetic atoms. 
The hybridization between the conduction and $f$-orbital on the magnetic atoms is represented by $V_{cf}$ and the $f$-electrons on the magnetic atoms 
experience the Coulomb repulsion of strength $U$. 
The first two lines on the right-hand side  of Eq.~(\ref{EQ:Hamil}) constitute the standard 
Anderson lattice model;  while the last two lines  represent the effect of a doped nonmagnetic atom, which without loss of generality 
is located at the origin $\mathbf{r}_I=(0,0)$.  The $f$-orbital energy level on the singly doped 
nonmagnetic atom is given by $\epsilon_{f}^{I}$.  In the Kondo hole 
problem, $\epsilon_{f}^{I}$ will be 
adjusted to ensure there is no $f$-electron occupation on the missing $f$-character center and 
as such the effect of on-site Coulomb repulsion can be negligible in the study of low energy 
electronic structure. 
In addition, the Kondo hole will also give rise to a potential scattering potential $\epsilon_{c}^{I}$ and a possible change of local hybridization $V_{cf}^{I}$. These three parameters make the description of a Kondo hole more realistic, demonstrating the flexibility of the Anderson lattice Hamiltonian. For simplicity, the effect of local hybridization change is neglected by assuming 
$V_{cf}^{I}=V_{cf}$ in the present work.

Due to the presence of onsite Hubbard interaction $U$ between the $f$-electrons on the magnetic atoms in Eq.~(\ref{EQ:Hamil}), the above problem is strongly correlated. This strong correlation
effect can be accounted for by reducing the statistical weight of double occupation in the Gutzwiller projected wavefunction approach~\cite{MCGutzwiller:1963}, and the projection can be carried out semi-analytically within the Gutzwiller approximation~\cite{MCGutzwiller:1965,DVollhardt:1984,FCZhang:1988}.  In the present problem, the lattice translation symmetry is broken
due to the Kondo hole, we use a spatially unrestricted Gutzwiller approximation (SUGA)~\cite{CLi:2006,JPJulien:2006,QHWang:2006,WHKo:2007,NFukushima:2008} to translate the original Hamiltonian Eq.~(\ref{EQ:Hamil}) into the following renormalized mean-field Hamiltonian:
\begin{eqnarray}
H_{\text{eff}}&=& - \sum_{ij,\sigma} (t_{ij}^{c} + \mu \delta_{ij})c_{i\sigma}^{\dagger} c_{j\sigma}  + \sum_{i\ne I ,\sigma}  [V_{cf} g_{i\sigma}  c_{i\sigma}^{\dagger} \tilde{f}_{i\sigma} + \text{H.c.}] \nonumber \\
&& + \sum_{i\ne I,\sigma} (\epsilon_{f}+\lambda_{i\sigma} -\mu) \tilde{f}_{i\sigma}^{\dagger} \tilde{f}_{i\sigma}
+ \sum_{i\ne I} U_{f} d_{i} +\sum_{\sigma} \epsilon_{c}^{I} c_{I\sigma}^{\dagger} c_{I\sigma} \nonumber \\
&& +   \sum_{\sigma} (\epsilon_{f}^{I} - \mu) \tilde{f}_{I\sigma}^{\dagger} \tilde{f}_{I\sigma}
+\sum_{\sigma}  [V_{cf}^{I}  c_{I\sigma}^{\dagger} \tilde{f}_{I\sigma} + \text{H.c.}]\;,
\label{EQ:Hamil_GA}
\end{eqnarray}  
where $\lambda_{i\sigma}$ and $d_{i}$ are the Lagrange multiplier and the double occupation 
at site $i$.  We have used $\tilde{f}_{i\sigma}^{\dagger}$ ($\tilde{f}_{i\sigma}$) to denote the quasiparticle field operators to differentiate from the truly $f$-electron operators in Eq.~(\ref{EQ:Hamil}). The local  $c$-$f$ hybridization has been renormalized by a factor of $g_{i\sigma}$,
which is given by 
\begin{equation}
g_{i\sigma} = \biggl{[} \frac{(\bar{n}_{i\sigma}^{\tilde{f}} -d_i)(1-  \bar{n}_{i}^{\tilde{f}}  + d_{i})}
{\bar{n}_{i\sigma}^{\tilde{f}}  (1-\bar{n}_{i\sigma}^{\tilde{f}} )}\biggr{]}^{1/2}
+ \biggl{[} \frac{d_i(\bar{n}_{i\bar{\sigma}}^{\tilde{f}}  - d_{i})}
{\bar{n}_{i\sigma}^{\tilde{f}}  (1-\bar{n}_{i\sigma}^{\tilde{f}} )}\biggr{]}^{1/2}\;,
\label{EQ:GF}
\end{equation} 
with $\bar{n}_{i\sigma}^{\tilde{f}}$ being the expectation value of the spin-$\sigma$ density operator $n_{i\sigma}^{\tilde{f}}=\tilde{f}_{i\sigma}^{\dagger} \tilde{f}_{i\sigma}$
and $\bar{n}_{i}^{\tilde{f}}=\sum_{\sigma} \bar{n}_{i\sigma}^{\tilde{f}}$.
Minimization of the expectation value of $H_{\text{eff}}$ leads to the following self-consistency 
conditions for $\lambda_{i\sigma}$ and $d_{i}$:
\begin{subequations}
\begin{eqnarray}
\lambda_{i\sigma} &=& V_{cf} \frac{\partial g_{i\sigma}}{\partial \bar{n}_{i\sigma}^{\tilde{f}}} 
(\langle c_{i\sigma}^{\dagger} \tilde{f}_{i\sigma}\rangle + \text{c.c}) \;, \\
-U &=& V_{cf} \sum_{\sigma} \frac{\partial g_{i\sigma}}{\partial d_{i}} 
(\langle c_{i\sigma}^{\dagger}\tilde{f}_{i\sigma}\rangle + \text{c.c})\;, 
\end{eqnarray}
\label{EQ:selfconsistency}
\end{subequations}
for $i \ne I$.  Equation~(\ref{EQ:Hamil_GA}) can be cast into the Anderson-Bogoliubov-de Gennes (Anderson-BdG) 
equations~\cite{JXZhu:2008}
\begin{equation}
\sum_{j} \left(
\begin{array}{cc}
h_{ij}^{c} &  \Delta_{ij}   \\
\Delta_{ji}^{*}    & h_{ij}^{\tilde{f}}
\end{array} \right)
\left( \begin{array}{c}
u_{j\sigma}^{n} \\
v_{j\sigma}^{n}
\end{array} \right)  = E_{n}
\left( \begin{array}{c}
u_{i\sigma}^{n} \\
v_{i\sigma}^{n}
\end{array} \right) \;,
\label{EQ:BdG}
\end{equation}
subject to the constraints given by Eq.~(\ref{EQ:selfconsistency}).
Here $h_{ij}^{c} = -t_{ij}^{c} -\mu \delta_{ij} + \epsilon_{I}^{c}\delta_{iI}\delta_{ij}$, 
$\Delta_{ij} =V_{cf} [g_{i} (1-\delta_{iI}) + \delta_{iI} ]\delta_{ij}$, 
 and $h_{ij}^{\tilde{f}} =  [ (\epsilon_{f}+\lambda_{i})(1-\delta_{iI}) + \epsilon_{f}^{I} \delta_{iI}  -\mu]
 \delta_{ij} $.   
Since most of heavy-fermion systems have a layered structure, 
in which the dominant effects occur in the planes containing 
$f$-electron atoms,  we solve this set of equations self-consistently via
exact diagonalization on a two-dimensional square lattice. 
After the self-consistency is achieved, one can then calculate the projected local density of states (LDOS) as defined by:
\begin{equation} 
(\rho_{i}^{c},\rho_{i}^{c\tilde{f}},\rho_{i}^{\tilde{f}}) = -2 \sum_{n} (\vert u_{i}^{n} \vert^2, u_{i}^{n}v_{i}^{n}, \vert v_{i}^{n} \vert^{2}) \frac{\partial f_{FD}(E-E_n)}{\partial E}\;,
\end{equation}
where the Fermi-Dirac distribution function $f_{FD}(E)=[\exp(E/k_{B}T)+1]^{-1}$.
Here we have used the fact that the eigenfunctions are real and the system has 
a two-fold spin degeneracy in the non-magnetic state, for which quantities like $\bar{n}_{i\sigma}^{\tilde{f}}$, $g_{i\sigma}$, and $\lambda_{i\sigma}$ become spin-independent.
Throughout this work, the quasiparticle
energy is measured with respect to the Fermi energy and the energy unit $t^c=1$
is chosen.

{\em Local electronic structure around the Kondo hole.~}
In our numerical calculations, we take the following values of parameters for the pristine system: The on-site Coulomb interaction on $f$-electrons $U=2$, the bare hybridization $V_{cf}=1$, 
and  the local $f$-level is taken to be $\epsilon_f = -1$. The temperature is fixed at $T=0.01$.
We solve the Anderson-BdG equations on a $32\times 32$ square lattice (by assuming a periodic boundary condition) to determine 
the self-consistency parameters $g_i$ and $\lambda_i$ while calculate the LDOS by using the supercell technique~\cite{JXZhu:1999} ($8 \times 8$ supercells are used). The chemical potential is varied so that the pristine system can be tuned into the HFL or KI  regime. To describe the Kondo hole, we have fixed the value of $\epsilon_{f}^{I}$ to be positive. This choice is consistent with the experimental realization of Kondo holes by substituting a Ce ion in a stoichiometric Ce compound by a La atom or a U ion in a U-heavy-fermion system by a Th atom, where the $f$-level is unoccupied by electrons on these impurity atoms. 

\begin{figure}[t!]
\centering\includegraphics[
width=1.0\linewidth,clip]{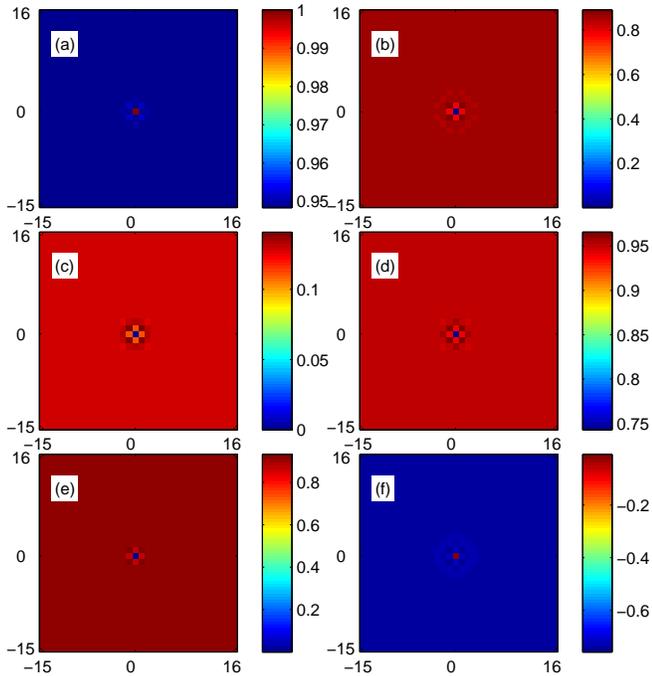}
\caption{(Color online)
Solution to the Kondo hole problem in the heavy Fermi liquid regime for $\mu=-0.4$. Contour plots of the Gutzwiller 
factor $g_{i}$ (a), the Lagrange multiplier $ \lambda_i $, the double occupation $d_i$ (c), the $c$-electron density $ \bar{n}_{i}^{c} $(d), $f$-electron density $ \bar{n}_{i}^{\tilde{f}} $ (e), 
and the hybridization density $\bar{n}_{i}^{c\tilde{f}}$ (f). The parameter values $\epsilon_{f}^{I}=100$ and 
$\epsilon_{c}^{I}=0$. The values of other parameters are given in the main text. 
 }
\label{fig:OP_HFL}
\end{figure}

Figure~\ref{fig:OP_HFL} shows the spatial variation of the self-consistently determined Gutzwiller factor $g_i$ (panel (a)), Lagrange multiplier $\lambda_{i}$ (panel (b)), the double occupation $d_i$, as well as partial charge density $\bar{n}_{i}^{c}$ (panel (d)), $\bar{n}_{i}^{\tilde{f}}$ (panel (e)),
and $\bar{n}_{i}^{c\tilde{f}}$ (panel (f)) around the Kondo hole in the HFL regime with $\mu=-0.4$. Here $\epsilon_{f}^{I}=100$ and $\epsilon_{c}^{I}=0$. The change of all these quantities happens around the Kondo hole.
The short length scale of the change is consistent with the coherence length as estimated by 
$\xi= \hbar \langle v_{F} \rangle/\pi g V_{cf}$, where $\langle v_F\rangle$ and $g$ are the averaged Fermi velocity and the Gutzwiller renormalization factor for the pristine system. With the given set of parameter values, we estimate $\hbar \langle v_F \rangle =4.89$, and  $g=0.95$ from the self-consistent iterations, which leads to $\xi$ being about 1.64 lattice constants. The anisotropy of the spatial change follows the underlying lattice, which is similar to that expected for a superconducting order parameter around a unitary non-magnetic impurity in a short-coherence $s$-wave superconductor. With the chosen parameter, it is found that the Kondo hole 
is negatively charged in the conduction band, by which we mean the conduction electron density on the Kondo hole is smaller than the value for the underlying pristine system.
To get a positively charged Kondo hole in the conduction band, an attractive potential scattering $\epsilon_{c}^{I}$ must be introduced.

\begin{figure}[b!]
\centering\includegraphics[
width=1.0\linewidth,clip]{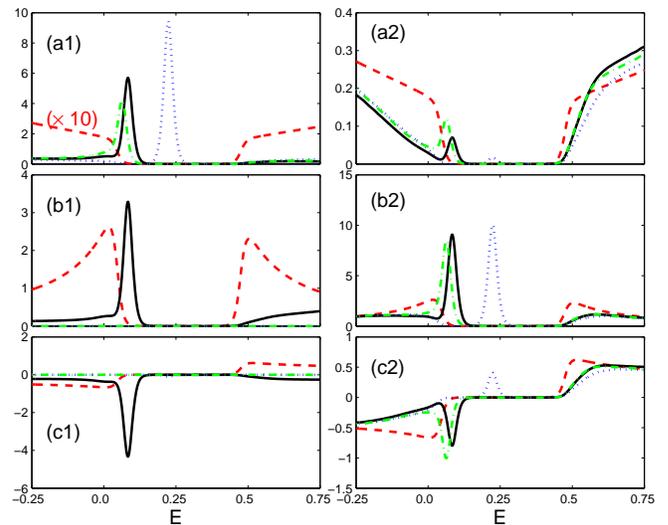}
\caption{(Color online)
Local density of states, in the HFL regime,  on the Kondo hole site (left column) and on its nearest neighboring site (right column).  Panels with label (aX), (bX), and (cX) correspond to the conduction band and $f$-quasiparticle, and hybrid density of states, respectively. The red-dashed line is for the pristine case; the black-solid and blue-dotted lines for $\epsilon_{f}^{I}=1.0$ and $100$ with  $\epsilon_{c}^{I}=0$ fixed, the green-dashdotted line for $\epsilon_{f}^{I}=100$
while $\epsilon_{c}^{I}=-1.0$. The coherent peaks for the pristine system are located at $E=0.016$ and $0.504$. The bound state peak  due to the presence of Kondo hole is located at $E=-0.082$ and $0.224$  for $\epsilon_{f}^{I}=1$ and $100$ with $\epsilon_{c}^{I}=0$ fixed, and at $E=0.064$ 
for $\epsilon_{f}^{I}=100$ but $\epsilon_{c}^{I}=-1.0$.
}
\label{fig:DOS_HFL}
\end{figure}

In Fig.~\ref{fig:DOS_HFL}, we show the local partial and hybrid DOS in the HFL regime 
 but with various values of $\epsilon_{f}^{I}$ and $\epsilon_{c}^{I}$ [The profile of self-consistency quantities around the Kondo hole remains similar to those shown in Fig.~\ref{fig:OP_HFL} and not shown here].  In the absence of Kondo hole, the LDOS is spatially independent, but both the local partial and hybrid DOS is asymmetric with respect to the chemical potential ($E=0$).
 The DOS intensity is peaked at the coherent gap edge (see the red-dashed line in the figure) and is finite at the Fermi energy, indicating the electron states are in the HFL regime. In the presence of Kondo hole, when 
 $\epsilon_{f}^{I}$ is increased,  an intragap bound state is formed with the energy position moved toward the center of the gap.
With increased $\epsilon_{f}^{I}$, the DOS intensity on the $f$-channel vanishes directly on the Kondo hole site while becomes stronger on the site nearest neighboring to the Kondo hole. However, the corresponding DOS in the conduction band has an opposite trend. Furthermore, an attractive potential scattering on the conduction channel will further tune the bound state peak away from the gap center.  It is also interesting to note that the hybrid DOS exhibits negative-sign peak for the intragap state for weak to intermediate values of $\epsilon_{f}^{I}$
 and $\epsilon_{c}^{I}$. 
Our result agrees with the STM observation of an electronic bound state
 at  thorium atoms, when they are substituted for U atoms in URu$_2$Si$_2$~\cite{MHHamidian:2011}.

\begin{figure}[t!]
\centering\includegraphics[
width=1.0\linewidth,clip]{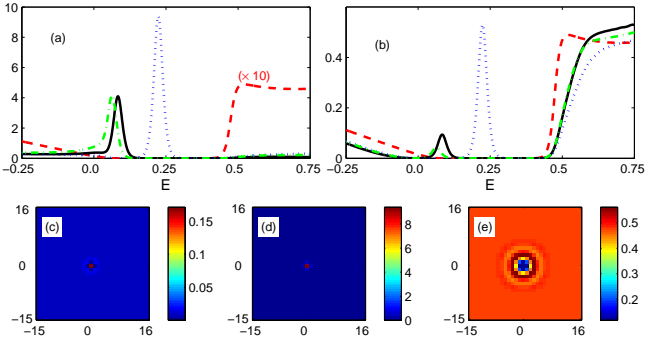}
\caption{(Color online) $dI/dV$, in the heavy Fermi liquid regime, as a function of energy on the Kondo hole (a) and its nearest neighboring site (b) corresponding to Fig.~\ref{fig:DOS_HFL}, 
as well as the $dI/dV$ spatial map with selected energies $E=0.016$ (c), 0.224 (d), and 0.504 (e). These energies correspond to the coherent peak position ($E= 0.016$ and $0.504$) and the intra-gap peak position ($E=0.224$) for $\epsilon_{f}^{I}=100$ and $\epsilon_{c}^{I}=0$.
}
\label{fig:didv_HFL}
\end{figure}

{\em Fano interference pattern around the Kondo hole.~}
To  model the STM electron tunneling,
for which a schematic picture can be found in Fig.~1 of Ref.~\onlinecite{MMaltseva:2009}, we introduce the tunneling Hamiltonian between the STM tip and the sample at a specified measure site ``$i$'': $H_{T}=\sum_{\sigma} V_{tc} \{c_{t\sigma}^{\dagger} [  c_{i\sigma} + (V_{tf}/V_{tc})  f_{i\sigma}] + \text{H.c.}\}$, where $V_{tc}$ and $V_{tf}$ are the amplitudes for tunneling into conduction and $f$-electron states.  Within the Gutzwiller approach to account for the $f$-electron correlation, tunneling amplitude $V_{tf}$ will be renormalized by the local Gutzwiller factor $g_{i}$, that is, $V_{tf} \rightarrow g_i V_{tf}$, which leads to,  in the weak-tunneling limit, the 
differential tunneling conductance~\cite{SupplMat}:
\begin{equation}
\frac{dI}{dV}= \frac{2e^2 \pi N_0 V_{tc}^2}{\hbar} [\rho_{i}^{c}(E) + 2r g_i \rho_{i}^{c\tilde{f}}(E) + r^{2} g_{i}^{2} \rho_{i}^{\tilde{f}}(E)]\;,
\label{EQ:didv}
\end{equation}
where $r=V_{tf}/V_{tc}$. Here we have assumed the tip density of states to be independent of energy and approximated its value at the Fermi energy $N_0$, which is on  the order of magnitude of  the inverse band width of the tip. Equation~(\ref{EQ:didv}) indicates that the renormalization due to 
the $f$-electron correlation effect in the heavy fermion materials must be taken into account explicitly in the tunneling conductance formula. This formula is logically consistent with the cotunneling mechanism~\cite{MMaltseva:2009}, and must be used in theoretical approaches
 based on an auxiliary field theory. 
 
As is shown in Fig.~\ref{fig:DOS_HFL}, the hybrid DOS is not positively definite, the line shape of $dI/dV$ is determined not only by whether the electronic structure for the pristine system has the particle-hole symmetry but also by the relative strength of tunneling amplitude of conduction band and $f$-electron channel.  Figure~\ref{fig:didv_HFL} shows the energy dependence of $dI/dV$ on the Kondo hole (panel (a)) and its nearest-neighboring site (panel (b)), as well as its spatial dependence 
 at selected energies (panels (c)-(e)), in the HFL regime with $r=0.2$. We used the 
conductance formula (Eq.~(\ref{EQ:didv})), for $T=0.01$ as a very low temperature solely for technical reasons. As can be seen from Fig.~\ref{fig:didv_HFL}(a)-(b), the Fano interference makes the $dI/dV$ characteristic strongly asymmetric, with the the continuum part of $dI/dV$ intensity at negative energies much smaller than at positive energies. The Fano interference also leads to different $dI/dV$ intensity maps at the two gap edges (see Fig.~\ref{fig:didv_HFL}(c) and (e)) with a ripple-like structure easily visible at the positive energy side. However, the peak structure due to the intragap bound state formed
 around the Kondo hole is robust against the Fano interference. On the Kondo hole site, the overall $dI/dV$ characteristic is similar to that of LDOS on the conduction band; while its nearest neighboring site,  the characteristic is sensitive to the detailed parameter values of $\epsilon_{f}^{I}$ and $\epsilon_{c}^{I}$. In addition, the spatial imaging of $dI/dV$ characteristic (Fig.~\ref{fig:didv_HFL}(d)) suggests the Kondo hole induced intragap states are localized states.

{\em Conclusion.}  We have studied  the local electronic structure around a single Kondo hole in an Anderson lattice model and the Fano interference pattern relevant to STM experiments.  Within the Gutzwiller method, we have obtained the existence of the intragap bound state induced around the Kondo hole in the  HFL regime.  The energy position of the intragap bound state is dependent on the on-site potential scattering  strength in the conduction and $f$-orbital channels. 
Within the same method, we have also derived a new $dI/dV$ formulation, which includes explicitly  the renormalization effect due to the $f$-electron correlation. It has been found that the Fano interference gives rise to highly asymmetric coherent peak structure separated by the hybridization gap.  The intragap peak structure has a Lorenzian shape. The corresponding $dI/dV$ intensity depends on the energy location of the bound state and the location of the STM measuring point. We have also addressed the Kondo hole problem in the Kondo insulator regime with $\mu=0$ (not shown).  The major finding for the HFL regime still holds except for the fact that the electronic states are fully gapped at the Fermi energy, and the Kondo hole becomes positively charged for $\epsilon_{c}^{I}=0$.  

This work was supported by U.S. DOE  at LANL  under Contract No. DE-AC52-06NA25396, 
U.S. DOE Office of Basic Energy Sciences
and in part by the Center for Integrated Nanotechnologies (J.-X.Z. and A.V.B.),  a U.S. DOE Office of Basic Energy Sciences user facility.  J.-X.Z. and A.V.B. also acknowledge partial support by the Aspen Center for  Physics under NSF Grant No. 1066293. J.-P.J. acknowledges the hospitality of LANL during his visits.

\end{document}


\title{Supplemental Material for Local Electronic Structure and Fano Interference in Tunneling into a Kondo Hole System}

\author{Jian-Xin Zhu}
\email[To whom correspondence should be addressed.\\]{jxzhu@lanl.gov}
\homepage{http://theory.lanl.gov}
\affiliation{Theoretical Division, Los Alamos National Laboratory,
Los Alamos, New Mexico 87545, USA}

\author{Jean-Pierre Julien}
\affiliation{Institut Neel CNRS and Universit\'{e} J. Fourier 25 Avenue des Martyrs, BP 166, F-38042
Grenoble Cedex 9, France}

\author{Y. Dubi}
\affiliation{School of Physics and Astronomy, Tel-Aviv University, Tel-Aviv, Israel }

\author{A. V. Balatsky}
\affiliation{Theoretical Division, Los Alamos National Laboratory,
Los Alamos, New Mexico 87545, USA}
\affiliation{Center for Integrated Nanotechnologies, Los Alamos National Laboratory, Los Alamos, New Mexico 87545, USA}

\maketitle

\onecolumngrid

\section{Derivation of the differential tunneling conductance within the Gutzwiller approximation}

In this supplemental material, we provide a more detailed discussion about the differential tunneling conductance within the Gutzwiller approximation. 

When the STM tip is positioned in the vicinity of site $i$, the tunneling Hamiltonian is given by: 
\begin{eqnarray}
H_{T} & = & \sum_{\sigma} [V_{tc}(c_{t\sigma}^{\dagger}c_{i\sigma} + \text{H.c.}) + 
V_{tf}(c_{t\sigma}^{\dagger} f_{i\sigma} + \text{H.c.})] \nonumber \\
&=& \sum_{\sigma} V_{tc} \{c_{t\sigma}^{\dagger} [  c_{i\sigma} + (V_{tf}/V_{tc})  f_{i\sigma}] + \text{H.c.}\} \;.
\end{eqnarray}
Here  $c_{t\sigma}^{\dagger}$ ($c_{t\sigma}$), $c_{i\sigma}^{\dagger}$ ($c_{i\sigma}$), $f_{i\sigma}^{\dagger}$ ($f_{i\sigma}$) are the creation (annihilation) operators for the STM tip electrons, and the conduction band electron and truly physical $f$-electrons at site $i$ with spin projection $\sigma$, while $V_{tc}$ and $V_{tf}$ are the amplitudes for tunneling into conduction and truly $f$-electron states. Therefore, the total Hamiltonian consists of three parts: 
\begin{equation}
H_{\text{total}} = H_{\text{tip}} + H_{\text{sample}} + H_{T}\;,
\end{equation}
with $H_{\text{tip}}= \sum_{\mathbf{k}\sigma} E_{t\mathbf{k}} c_{t\mathbf{k}\sigma}^{\dagger} c_{t\mathbf{k}\sigma}$ represents the Hamiltonian for the STM tip while $H_{\text{sample}}$ represents the sample system (that is the heavy fermion system) as  given by Eq.~(1) in the main text. The electron tunneling operator is defined as the rate of change of the tip electrons $\hat{n}_{t}=\sum_{\sigma} c_{t\sigma}^{\dagger} c_{t\sigma}$:
\begin{eqnarray}
\hat{I} &=& \frac{ie}{\hbar}[\hat{n}_{t}, H_{\text{total}}]_{-} \nonumber \\
&=& \frac{ie}{\hbar} \sum_{\sigma} V_{tc} \{c_{t\sigma}^{\dagger} [  c_{i\sigma} + (V_{tf}/V_{tc})  f_{i\sigma}] - \text{H.c.}\} \;.
\end{eqnarray}
where $e$ is the electron charge and $\hbar$ is the reduced Planck constant. Within the linear response theory, the expectation value of $\hat{I}$ is derived as follows~\cite{MahanBook}:
\begin{eqnarray}
\langle \hat{I} \rangle & = & -\frac{e}{\hbar} \int_{-\infty}^{t} dt^{\prime} 
 \{ V_{tc}^{2}  \langle [\hat{A}(t), \hat{A}^{\dagger}(t^{\prime})]_{-}\rangle e^{-ieV(t-t^{\prime})} 
- V_{tc}^{2} \langle [\hat{A}^{\dagger}(t),\hat{A}(t^{\prime})]_{-} \rangle e^{ieV(t-t^{\prime})}
 \nonumber \\ 
 &&+V_{tc} V_{tf}  \langle [\hat{A}(t), \hat{B}^{\dagger}(t^{\prime})]_{-}\rangle e^{-ieV(t-t^{\prime})} 
- V_{tc} V_{tf} \langle [\hat{A}^{\dagger}(t),\hat{B}(t^{\prime})]_{-} \rangle e^{ieV(t-t^{\prime})} \nonumber \\
&&+V_{tc} V_{tf}  \langle [\hat{B}(t), \hat{A}^{\dagger}(t^{\prime})]_{-}\rangle e^{-ieV(t-t^{\prime})} 
- V_{tc} V_{tf} \langle [\hat{B}^{\dagger}(t),\hat{A}(t^{\prime})]_{-} \rangle e^{ieV(t-t^{\prime})} \nonumber \\
&& + V_{tf}^{2}  \langle [\hat{B}(t), \hat{B}^{\dagger}(t^{\prime})]_{-}\rangle e^{-ieV(t-t^{\prime})} 
- V_{tf}^{2} \langle [\hat{B}^{\dagger}(t),\hat{B}(t^{\prime})]_{-} \rangle e^{ieV(t-t^{\prime})}   \} \nonumber \\ 
&=& -\frac{ie}{\hbar} \biggl{\{} V_{tc}^{2} [U_{\text{ret}}^{AA}(-eV) - U_{\text{adv}}^{AA}(-eV)]  + V_{tf}^{2} [U_{\text{ret}}^{AA}(-eV) - U_{\text{adv}}^{AA}(-eV)] \nonumber \\
&& + V_{tc}V_{tf} [U_{\text{ret}}^{AB}(-eV) - U_{\text{adv}}^{AB}(-eV)] 
  + V_{tc}V_{tf} [U_{\text{ret}}^{BA}(-eV) - U_{\text{adv}}^{BA}(-eV)] \biggr{\}}  \nonumber \\
 &=& \frac{2e}{\hbar} \{V_{tc}^{2} \text{Im}[U_{\text{ret}}^{AA}(-eV)] + V_{tc}V_{tf} \text{Im}[U_{\text{ret}}^{AB}(-eV)] + V_{tc} V_{tf} \text{Im}[U_{\text{ret}}^{BA}(-eV)] + V_{tf}^{2} \text{Im}[U_{\text{ret}}^{BB}(-eV)]\}  \;.
\end{eqnarray}
Here $\hat{A}(t) =  \sum_{\mathbf{k}\sigma}c_{i\sigma}^{\dagger}(t) c_{t\mathbf{k}\sigma}(t)/\sqrt{\mathcal{N}_{\text{tip}}}$ and $\hat{B}(t) =  \sum_{\mathbf{k}\sigma}  
f_{t\mathbf{k}\sigma}^{\dagger}(t)c_{t\mathbf{k}\sigma}(t) /\sqrt{\mathcal{N}_{\text{tip}}}$, where $\mathcal{N}_{\text{tip}}$ is the normalization factor for the tip electron degrees of freedom and $c_{i\sigma}^{\dagger} (t)= e^{iK_{\text{sample}}t}c_{i\sigma}^{\dagger}  e^{-iK_{\text{sample}}t}$, $f_{i\sigma}^{\dagger} (t)= e^{iK_{\text{sample}}t} f_{i\sigma}^{\dagger}  e^{-iK_{\text{sample}}t}$, and $c_{t\sigma} (t)= e^{iK_{\text{tip}}t}c_{t\sigma} e^{-iK_{\text{tip}}t}$. The symbols $K_{\text{sample}}$ and $K_{\text{tip}}$ denote the Hamiltonian with respect to the respective chemical potentials: $K_{\text{sample}} = H_{\text{sample}} - \mu_{\text{sample}} N_{\text{sample}}$ and $K_{\text{tip}}= H_{\text{tip}} - \mu_{\text{tip}} N_{\text{tip}}$, where $N_{\text{sample}}$ and $N_{\text{tip}}$ are the total electron numbers in the sample and the tip, respectively. The chemical potential difference between the tip and the sample is identified as the voltage bias $\mu_{\text{tip}} - \mu_{\text{sample}}= eV$. The retarded correlations are defined as:
\begin{subequations}
\begin{eqnarray}
U_{\text{ret}}^{AA}(t) &=& -i\theta(t) \langle [\hat{A}(t), \hat{A}^{\dagger}(0)]_{-}\rangle  \;,\\ 
U_{\text{ret}}^{AB}(t) &=& -i\theta(t) \langle [\hat{A}(t), \hat{B}^{\dagger}(0)]_{-}\rangle  \;,\\
U_{\text{ret}}^{BA}(t) &=& -i\theta(t) \langle [\hat{B}(t), \hat{A}^{\dagger}(0)]_{-}\rangle  \;,\\
U_{\text{ret}}^{BB}(t) &=& -i\theta(t) \langle [\hat{B}(t), \hat{B}^{\dagger}(0)]_{-}\rangle  \;.
\end{eqnarray}
\end{subequations}
The Fourier transform of these retarded correlation can be evaluated more easily from their corresponding Matsubara correlation functions, which are found to be:
\begin{subequations}
\begin{eqnarray}
U^{AA}(i\omega_m) = \frac{1}{\mathcal{N}_{\text{tip}}} \sum_{\mathbf{k}\sigma} \frac{1}{\beta} \sum_{ip_n} G_{i\sigma}^{cc}(ip_{n}) G_{t,\mathbf{k}\sigma}(ip_{n}+i\omega_{m})\;,\\
U^{AB}(i\omega_m) = \frac{1}{\mathcal{N}_{\text{tip}}} \sum_{\mathbf{k}\sigma} \frac{1}{\beta} \sum_{ip_n} G_{i\sigma}^{fc}(ip_{n}) G_{t,\mathbf{k}\sigma}(ip_{n}+i\omega_{m})\;, \\
U^{BA}(i\omega_m) = \frac{1}{\mathcal{N}_{\text{tip}}} \sum_{\mathbf{k}\sigma} \frac{1}{\beta} \sum_{ip_n} G_{i\sigma}^{cf}(ip_{n}) G_{t,\mathbf{k}\sigma}(ip_{n}+i\omega_{m})\;, \\
U^{BB}(i\omega_m) = \frac{1}{\mathcal{N}_{\text{tip}}} \sum_{\mathbf{k}\sigma} \frac{1}{\beta} \sum_{ip_n} G_{i\sigma}^{ff}(ip_{n}) G_{t,\mathbf{k}\sigma}(ip_{n}+i\omega_{m})\;,
\end{eqnarray}
\end{subequations}
where $\beta=1/k_{B}T$ with $k_B$ and $T$ the Boltzmann constant and the electron temperature, respectively, while $p_n = (2n\pi+1)k_{B}T$ and $\omega_m = 2m\pi k_{B}T$ are the Matsubara frequency for fermions and bosons, respectively. The electron Green's functions for different parts can be written as:
\begin{subequations}
\begin{eqnarray}
G_{t,\mathbf{k}\sigma}(ip_n + i\omega_m)  &=&  \frac{1}{ip_n + i\omega_m -\xi_{\mathbf{k}}} \;, \\
G_{i\sigma}^{cc}(ip_n) &=& \int_{-\infty}^{\infty} d\epsilon \frac{\rho_{i\sigma}^{c}(\epsilon)}{ip_n-\epsilon}\;, \\
G_{i\sigma}^{cf(fc)}(ip_n) &=& \int_{-\infty}^{\infty} d\epsilon \frac{\rho_{i\sigma}^{cf(fc)}(\epsilon)}{ip_n-\epsilon}\;, \\
G_{i\sigma}^{ff}(ip_n) &=& \int_{-\infty}^{\infty} d\epsilon \frac{\rho_{i\sigma}^{f}(\epsilon)}{ip_n-\epsilon}\;, 
\end{eqnarray}
\end{subequations}
where we have introduced the local single-particle and hybridization density of states of truly {\em physical electrons}: $ \rho_{i\sigma}^{c}(\epsilon)$, $\rho_{i\sigma}^{f}(\epsilon)$, and $\rho_{i\sigma}^{cf(fc)}(\epsilon)$. Performing the frequency summation, we obtain the current as:
\begin{equation}
\langle \hat{I} \rangle = -\frac{2e\pi}{\hbar \mathcal{N}_{\text{tip}}} \sum_{\mathbf{k} \sigma}   \int d\epsilon [V_{tc}^{2} \rho_{i\sigma}^{c}(\epsilon) + 2V_{tc} V_{tf} \rho_{i\sigma}^{cf}(\epsilon) + V_{tf}^{2} \rho_{i\sigma}^{f}(\epsilon)] [f_{\text{FD}}(\epsilon) - f_{\text{FD}}(\xi_{\mathbf{k}})] \delta(-eV + \epsilon -\xi_{\mathbf{k}})\;.
\end{equation}
Within the wide-band approximation for the tip electrons, we arrive at 
\begin{equation}
\langle \hat{I} \rangle = -\frac{2e\pi N_0}{\hbar}  \sum_{\sigma}   \int d\epsilon [V_{tc}^{2} \rho_{i\sigma}^{c}(\epsilon) + 2V_{tc} V_{tf} \rho_{i\sigma}^{cf}(\epsilon) + V_{tf}^{2} \rho_{i\sigma}^{f}(\epsilon)] [f_{\text{FD}}(\epsilon) - f_{\text{FD}}(\epsilon-eV)] \;.
\end{equation}
where $N_0$ is the density of states for the STM tip.
The differential tunneling conductance is then found to be:
\begin{equation}
\frac{d\langle I \rangle}{dV} = \frac{2e^2 \pi N_0}{\hbar} \sum_{\sigma} \int d\epsilon 
 [V_{tc}^{2} \rho_{i\sigma}^{c}(\epsilon) + 2V_{tc} V_{tf} \rho_{i\sigma}^{cf}(\epsilon) + V_{tf}^{2} \rho_{i\sigma}^{f}(\epsilon)] \biggl{(} -\frac{\partial f_{\text{FD}}(\epsilon-eV)}{\partial \epsilon}\biggr{)}\;,
 \end{equation}
 and has the zero-temperature limit
\begin{equation}
\frac{d\langle I \rangle}{dV} = \frac{2e^2 \pi N_0}{\hbar} \sum_{\sigma}  
 [V_{tc}^{2} \rho_{i\sigma}^{c}(eV) + 2V_{tc} V_{tf} \rho_{i\sigma}^{cf}(eV) + V_{tf}^{2} \rho_{i\sigma}^{f}(eV)] \;.
 \end{equation} 
The same type of formula can be written for a finite temperature when the thermalized local density of states is defined as in Eq.~(6) of the main text. The remaining task is to evaluate the truly electron local density of states in the sample. It is a nontrivial task in strongly correlated electron systems and except for a few rare cases approximations are usually made. In the methods like non-crossing approximation (NCA)~\cite{NEBickers:1986,ASchiller:2000}, equation-of-motion (EOM) decoupling scheme~\cite{CLacroix:1981,YQi:2009}, numerical renormalization group (NRG) method~\cite{RBulla:2008}, or quantum Monte Carlo (QMC) simulation~\cite{MJarrell:1996}, the electron GreenÕs functions are directly evaluated. In this fashion, the renormalization from correlation effects is directly encoded into the electron self-energy and the above tunneling conductance can be applied directly. However, when we solve the strongly correlated electron systems within an auxiliary field theory, a caution must be taken. In the Gutzwiller approximation, we note that the coherent part of  the Green's functions involving the truly $f$-electrons are related to the Green's functions involving the quasiparticle operator as $G_{i\sigma}^{cf}(\omega) = g_{i\sigma} G_{i\sigma}^{c\tilde{f}}(\omega)$ and $G_{i\sigma}^{ff}(\omega) = g_{i\sigma}^{2}G_{i\sigma}^{\tilde{f}\tilde{f}}(\omega)$, where $G_{i\sigma}^{c\tilde{f}}(\omega)$ and $G_{i\sigma}^{\tilde{f}\tilde{f}}(\omega)$ are determined by the solutions to the effective Hamiltonian given by Eq.~(2) in the main text. With these relations, the zero-temperature differential tunneling conductance reduces to Eq.~(7) in the main text. The same type of formula has also been derived in the large-$N$ mean-field approach to the Kondo lattice~\cite{MMaltseva:2009}. We conclude that it should be applicable to all theoretical approaches based on an auxiliary field theory (like slave-boson mean-field theory~\cite{GKotliar:1986}).